\documentclass{aastex631}

\defcitealias{Porquet2000}{PD00}
\defcitealias{GW_Note1}{Note I}

\shorttitle{Study of the Soft X-ray Emission Lines in NGC 4151}
\shortauthors{Grafton-Waters et al.}

\begin{document}

	\title{A Study of the Soft X-ray Emission Lines in NGC 4151 \\
		II. The Internal Plasma Properties}
	%\maketitle
	
	%\correspondingauthor{S. Grafton-Waters}
	%\email{sam.waters.17@ucl.ac.uk}
	\author[0000-0002-4833-8612]{S. Grafton-Waters}
	\author[0000-0002-0383-6917]{W. Dunn}
	\affiliation{Mullard Space Science Laboratory, University College London, Holmbury St. Mary, Dorking, Surrey, RH5 6NT, UK}
	
	\begin{abstract}
		We showcase a tool suite that enables the fitting of soft X-ray spectra in active galactic nuclei (AGN), without the need for specialist software, allowing access to AGN physics for school students. While these standardised Python tools were useful for measuring velocities (\citetalias{GW_Note1}), they offered significantly fewer capabilities for radiative recombination continua (RRC), and R and G ratios, utilised to obtain the internal plasma properties within the outflowing wind seen in NGC 4151. Although further work is required for these tools to be used in outreach projects, we present findings of the plasma temperature and density in NGC 4151 spanning a 15 year period. 
	\end{abstract}
	
	\section{Introduction}
	The outflowing winds seen in many active galactic nuclei (AGN) are photoionised (PI) by the radiation from the central source \citep[e.g.][]{Grafton-Waters2020}. The signatures in the soft X-ray spectra of these winds are the narrow absorption lines from the warm absorber \citep[e.g.][]{Blustin2005} and strong emission lines from the narrow line region \citep[NLR; e.g.][]{Grafton-Waters2021}. 
	
	A useful diagnostic tool to determine the internal properties of the plasma are the radiative recombination continua (RRC) features, produced when free electrons recombine to the ground state of an ion. For PI plasma, the RRC is narrow and strong, whereas for collisionally ionised (CI) plasma, the line is broad and weak. The width ($\Delta E$) of the RRC is approximately equal to the thermal energy of the free electrons in the plasma, $kT$, where \textit{k} is the Boltzmann constant and $T$ is the electron temperature \citep{Kahn2002}. The temperatures are of order $T \sim 10^4$ K for PI plasmas and $T \geq 5 \times 10^5$ for CI plasmas \citep[e.g.][hereafter PD00]{Porquet2000}. 
	
	Another essential tool for identifying plasma properties are the R and G ratios (\citetalias{Porquet2000}), defined as $R = f/i$ and $G = (f + i) / r$, where r, i, f are the fluxes of the resonance, intercombination and forbidden lines of the He-like triplets, respectively \citep[see Figure 1 in][hereafter Note I]{GW_Note1}. These ratios are used for density and temperature diagnostics, respectively. However, R and G ratios often lead to contradictory and inconsistent measurements of the plasma conditions, leaving the temperature and density of the plasmas inconclusively identified \citep[e.g][]{Kinkhabwala2002, Schurch_RGS}.
	
	%\textbf{The ratios for the O VII and N VI triplet lines in each epoch were calculated using the normalisation from each line, found from the Gaussian model \citep[][hereafter Note I]{GW_Note1}.}
	
	In \citetalias{GW_Note1}, we showcased a tool suite that enabled the fitting of the XMM-Newton reflection grating spectrometer \citep[RGS;][]{denHerder2001} spectra from NGC 4151 to identify previously unreported plasma regions. This new toolkit did not require specialist software \citep[e.g. SPEX;][]{Kaastra2017} which may sometimes inhibit access to science by e.g. school students. Here, we expand this toolkit to conduct modelling techniques of the internal plasma properties in NGC 4151, and to test whether studies of more complex AGN features within an outreach project can be achieved. %This will give constraints on the ionised plasma without the need for complex photoionisation models. 
	The code can be found at \cite{sgwxray_2021_5116838}. 
	
	The aim of this Note is to determine the internal properties of the outflowing wind in NGC 4151. The RGS spectra were reduced as in \citetalias{GW_Note1}, except that all observations in the same epoch were combined using \texttt{RGSCOMBINE}. The six epochs are presented in Figure \ref{Fig:Figure}. % The RRC features in the RGS spectra from the six epochs are fitted, in addition to determining the line ratios of the O VII and N VI triplet lines to estimate the densities. 

	%\textbf{R and G give T and n diagnositics, not the `type' of plasma}

	\section{RRC Features}
	
	%through \texttt{RGSPROC} command in the SAS software \texttt{v17.0.0}\footnote{See \url{https://www.cosmos.esa.int/web/xmm-newton/sas-thread-rgs}}. Any large background counts in CCD9 were removed before combining all RGS spectra for all observations within the epoch using \texttt{RGSCOMBINE}. 
	
\begin{figure*}
		\centering
		\epsscale{1.15}
		\plotone{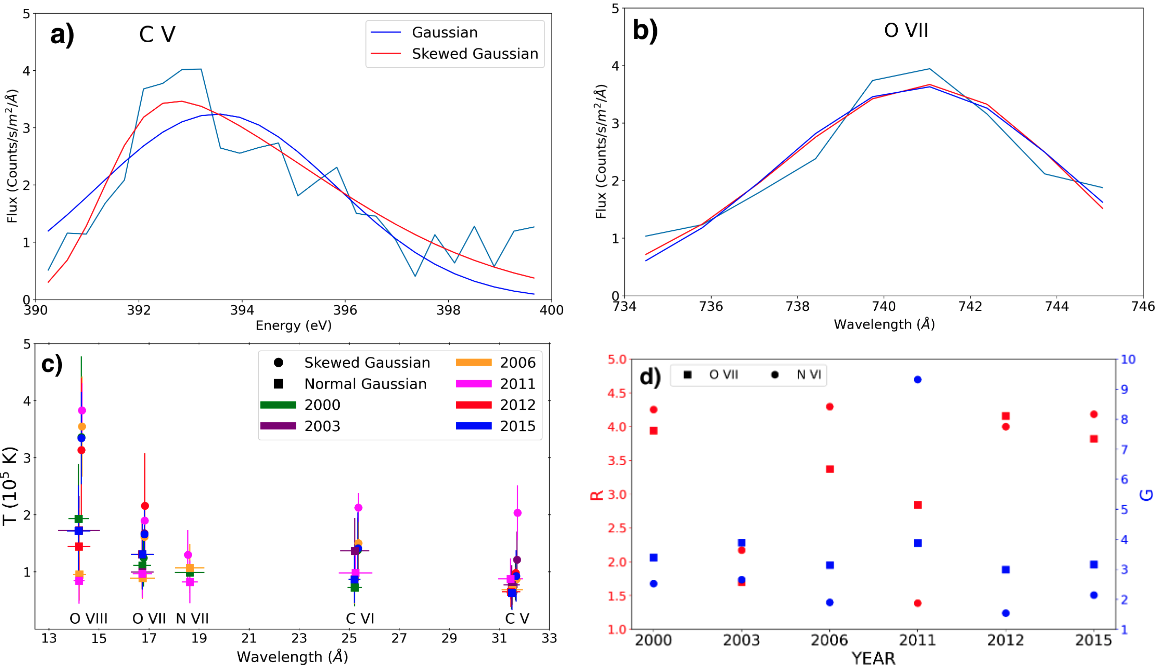}
		\caption{RRC features of C V \textbf{(a)} and O VII \textbf{(b)}. The C V RRC has the characteristic thermal distribution shape, while the O VII RRC does not. Both these features (light-blue lines) were fitted with the SGM (red line) and NGM (dark-blue line); there is negligible difference between models in \textbf{(b)}. \textbf{(c)} Temperatures of each RRC feature as measured from the modelling; the colours represent one of the six epochs studied. \textbf{(d)} R (red) and G (blue) ratios for the O VII (squares) and N VI (circles) triplet lines from each epoch.}
		\label{Fig:Figure}
	\end{figure*}

	In each spectrum, the five strongest RRC lines were O VIII (14.2 \AA), O VII (16.8 \AA), N VII (18.6 \AA), C VI (25.4 \AA), and C V (31.6 \AA). Initially, we modelled these features with a Maxwell thermal distribution model in \texttt{LMFIT}\footnote{\url{https://lmfit.github.io/lmfit-py/builtin_models.html\#thermaldistributionmodel}} \citep{LMFIT2014}, but this did not work, either because the energy range was too small or not all the features showed the classical Maxwellian profile (see Figures \ref{Fig:Figure} (a) and (b)). Instead, we successfully modelled the RRCs with a skewed-Gaussian model (SGM\footnote{\url{https://lmfit.github.io/lmfit-py/builtin_models.html\#skewedgaussianmodel}}), given by
	\begin{equation}
	F(x) = \frac{A}{\sqrt{2\pi \sigma^2}}\exp{\left[-\frac{(x - \mu)^2}{2\sigma^2}\right]} \left(1 + erf\left[ \frac{\gamma (x - \mu)}{\sqrt{2\sigma^2}} \right] \right),
	\label{EQ:Skew_Gaus}
	\end{equation}
	where $\gamma$ is the skewness parameter and $erf$ is the error function\footnote{See Section 4.2.1 in \url{https://dyedavid.files.wordpress.com/2016/11/mse101.pdf}}, defined as 
	\begin{equation}
	erf(x) = \frac{1}{\sqrt{\pi}} \int_{x_1}^{x_2} e^{-t^{2}} dt.
	\end{equation} 

However, in pursuing a SGM we lose the ability to interpret the true plasma temperature from the RRC width, $\Delta E$. In order to utilise this tool in future outreach projects this problem needs solving. Here, we measure $\Delta E = 2.355 \sigma$, where $\sigma$ is the standard deviation of SGM, to test how far our values are from expectation. In addition, we compared results to using $\Delta E = \sigma \frac{\sinh \gamma}{\gamma}$ \citep{Rusch1973}, but this gave infeasible temperatures ($T > 1 \times 10^6$ K) which are therefore not plotted in Figure \ref{Fig:Figure} (c). Furthermore, some RRC features appeared to be similar to the normal emission lines in the rest of the spectrum (Figure \ref{Fig:Figure} (b)). Therefore, the RRCs were also fitted with a normal Gaussian model (NGM), where $\Delta E = 2.355 \sigma$, by definition.

	Figure \ref{Fig:Figure} (c) compares temperatures from the RRCs fitted in each epoch with SGM and NGM. The temperatures of these RRC features are $T \sim 1 - 4 \times 10^5$ K, an order of magnitude larger than those ($T = 2 - 7 \times 10^4$ K) found by \cite{Schurch_RGS}. Neither model is consistent with PI plasma because we are unable to obtain the true $\Delta E$ of the RRC features. Usually in photoionisation modelling, the plasma temperature is measured through ionisation and energy balance calculations and line ratios; here the temperature is instead estimated using an incorrect line widths because of the limitations of SGM or NGM. 
	
	%Therefore, although the temperatures in Figure \ref{Fig:Figure} a) are significantly larger compared to the values found by \cite{Schurch_RGS}, they can still be used as an indicator about the properties of the plasma. 

	%\textbf{From Figure \ref{Fig:Figure} b) the 2011 O VII R ratio is the only value to be consistent with the expected value. In all the other epochs the values are too high, whereas for the N VI ratios are much lower than their expected values. For the G ratios (right side of Figure \ref{Fig:4151_Ratios}) the expected values are the same for both triplets. The O VII ratios are closer to the expected G ratio (2012 is the only value within the range) compared to the N VI ratios. \textbf{What does this mean?} This is similar to the findings in NGC 1068 by \cite{Kinkhabwala2002}. They found the G ratios to be in between CIE and PIE plasmas for both ions and the observed R ratio for O VII was too high. The R ratio for N VI was the only value to be consistent with PIE plasma. 

	\section{R and G Ratios}
	
	%, the G and R ratios were used to estimate the temperature and density of the plasma regions.
	The R and G ratios were measured using fluxes of the r, i and f lines from the O VII and N VI triplet lines. We then used Figures 7 and 8 from \citetalias{Porquet2000} to obtain temperatures and densities ($n_e$). From Figure \ref{Fig:Figure} (d), $G_{N VI} = 1 - 3$ and $G_{O VII} = 2 - 4$; $R_{N VI} = 1 - 5$ and $R_{O VII} = 3 - 5$. These G ratios correspond to $T \simeq 1 - 2 \times 10^6$ K, using Figure 7 from \citetalias{Porquet2000}, while the R ratios give $n_e \simeq 10^{15} - 5\times 10^{16}$ m\textsuperscript{-3} for N VI and $n_e \leq 10^{15}$ m\textsuperscript{-3} for O VII when using Figure 8 from \citetalias{Porquet2000}, with the $T \sim 2 \times 10^6$ K lines.
	
	Interestingly, temperatures found from the G ratios are an order of magnitude larger than the values from the RRC modelling. This comes about because the G values obtained here are between 1 and 4, whereas for pure PI plasmas, the expected G ratios are between 4.0 and 5.5 \citep{Kinkhabwala2002}. The densities from the R ratios imply that the plasma originates from the broad line region (BLR), but this is unlikely given that in \citetalias{GW_Note1} we obtained distances consistent with the NLR ($R \geq 10$ pc). \cite{Schurch_RGS} calculated the R and G ratios for Ne IX, O VII, N VI (their Table 3), finding similar results to those here. % Alternatively, PI modelling in spectral codes have updated line ratios and energy balance equations \citep{Mehdipour2016}, such that they are significantly improved compared to the using the Figures in \citetalias{Porquet2000}.
	
	%\textbf{Find T and n from R and G plots in Porquet+2000. Compare with the RRC temperatures. \\Gaussian or skewed Gaussian? \\ Are the line widths related to the RGS resolution at lower wavelengths? }

	\section{Conclusion}
	
	For NGC 4151, the temperatures from the RRC lines and G ratios were $T \sim 10^5 - 10^6$K, which are inconclusive and inconsistent for either PI or CI plasmas, but do agree with previous measurements \citep{Schurch_RGS}. Meanwhile, R ratios suggest the emitting plasma regions have a density between $n_e \sim 10^{15} - 10^{16}$ m\textsuperscript{-3}, implying the plasma is located in the BLR, not the NLR. %This is consistent with the temperature values, but not with a PI plasma.
	
	From this study, it is clear that while this method of analysing the emission lines of AGN provides reasonable velocities of the plasma regions, density and temperature diagnostics through RRC modelling and R and G ratios are poor. Although PI modelling with spectral codes is the only way to obtain plasma measurements, this work proves we are able to obtain upper estimates of the internal plasma properties of AGN winds using standardised python tools. Therefore, AGN science can be executed in public engagement projects with schools that cannot access specialist spectral analysis codes. 
	
%	\begin{acknowledgments}
%		S.G.W. acknowledges the support of a PhD studentship awarded by the UK Science \& Technology Facilities Council (STFC).
%	\end{acknowledgments}

%	\bibliography{references.bib}
%	\bibliographystyle{aasjournal}
%	
\end{document}